\documentclass[aps,prd,groupedaddress,twocolumn,showpacs,nofootinbib]{revtex4}
\usepackage{graphicx,epsfig}

\begin{document}


\title{Generalised constraints on the curvature perturbation from primordial
black holes}



\author{Amandeep S. Josan}
\email[]{ppxaj1@nottingham.ac.uk}


\affiliation{School of Physics and Astronomy, University of Nottingham, University Park, Nottingham, NG7 2RD, UK}

\author{Anne M. Green}
\email[]{anne.green@nottingham.ac.uk}


\affiliation{School of Physics and Astronomy, University of Nottingham, University Park, Nottingham, NG7 2RD, UK}

\author{Karim A. Malik}
\email[]{k.malik@qmul.ac.uk}


\affiliation{Astronomy Unit, School of Mathematical Sciences, Queen Mary, University of
London, Mile End Road, London E1 4NS, UK}


\date{\today}

\begin{abstract}

Primordial black holes (PBHs) can form in the early Universe via the
collapse of large density perturbations. There are tight constraints
on the abundance of PBHs formed due to their gravitational effects
and the consequences of their evaporation.  These abundance
constraints can be used to constrain the primordial power spectrum,
and hence models of inflation, on scales far smaller than those
probed by cosmological observations.  We compile, and where relevant
update, the constraints on the abundance of PBHs before calculating
the constraints on the curvature perturbation, taking into account
the growth of density perturbations prior to horizon entry. We
consider two simple parameterizations of the curvature perturbation
spectrum on the scale of interest: constant and power-law.  The
constraints from PBHs on the amplitude of the power spectrum are
typically in the range $10^{-2}-10^{-1}$ with some scale dependence.

\end{abstract}

\pacs{98.80.Cq}

\maketitle


\section{Introduction}


Primordial black holes (PBHs) can form in the early Universe via the
collapse of large density
perturbations~\cite{Carr:1974nx,Carr:1975qj}. If the density
perturbation at horizon entry in a given region exceeds a threshold
value, of order unity, then gravity overcomes pressure forces and the
region collapses to form a PBH with mass of order the horizon mass.
There are a number of limits, spanning a wide range of masses, on the
PBH abundance. PBHs with mass $M_{\rm PBH} \lesssim 5 \times 10^{14}
\, {\rm g}$ will have evaporated by the present
day~\cite{Hawking:1974sw,MacGibbon:2007yq} and their abundance is
constrained by the consequences of the Hawking radiation emitted.
More massive PBHs are constrained by their present day gravitational
effects. The resulting limits on the initial mass fraction of PBHs are
very tight, $\beta \equiv \rho_{\rm PBH}/\rho_{\rm tot} < {\cal O}
(10^{-20} )$, and can be used to constrain the power spectrum of the
primordial density, or curvature, perturbations (see
e.g. Ref.~\cite{Carr:2005zd}).

The power spectrum of the primordial curvature perturbation,
${\cal{P}}_{\cal{R}}(k)$, on cosmological scales is now accurately
measured by observations of the cosmic microwave background
(CMB)~\cite{Dunkley:2008ie} and large scale
structure~\cite{Cole:2005sx,Tegmark:2006az}. These measurements can be
used to constrain, and in some cases exclude, inflation models
(c.f. Ref.~\cite{Peiris:2006sj}). Cosmological observations span a
relatively small range of scales (comoving wavenumbers between 
$k\sim 1 \, {\rm Mpc}^{-1}$ and $k\sim 10^{-3} \, {\rm Mpc}^{-1}$), 
and hence probe a limited region of the inflaton potential. The PBH 
constraints on the curvature power spectrum are fairly weak; the upper 
limit is many orders of magnitude larger than the measurements on cosmological 
scales. They do, however, apply over a very wide range of scales (from 
$k\sim 10^{-2} \, {\rm Mpc}^{-1}$ to $k\sim 10^{23} \, {\rm Mpc}^{-1}$) 
and therefore provide a useful constraint on models of inflation~\cite{Carr:1993aq}.

The simplest assumption for the power spectrum is a scale-free
power law with constant spectral index, $n$:
\begin{equation} 
\label{powerlawps}
{\cal{P}}_{\cal{R}} (k) \equiv \frac{k^3}{2 \pi^2} \langle
        |{\cal{R}}_{k}|^2 \rangle  = {\cal{P}}_{\cal R}(k_0) 
      \left( \frac{k}{k_{0}} \right)^{n-1} \,,
\end{equation}
where $k_{0}$ is a suitably chosen normalisation scale.  In this case the 
PBH abundance constraints require $n<1.25-1.30$~\cite{Carr:1994ar,Green:1997sz,Kim:1996hr,Bringmann:2001yp}.
The spectral index on cosmological scales is, however, now accurately
measured: $n = 0.963^{+0.014}_{-0.015} $~\cite{Dunkley:2008ie}. In
other words, if the power spectrum is a pure power law then the number
of PBHs formed will be completely negligible. However, as we will now
outline, if the primordial perturbations are produced by inflation
then the power spectrum is not expected to be an exact power law over
all scales.

The power spectrum produced by slow-roll inflation can be
written as an expansion about a wave number $k_{0}$
(e.g. Ref.~\cite{Lidsey:1995np})
\begin{eqnarray}
\label{powernalpha}
\ln{{\cal P}_{\cal R} (k)} &\approx& \ln{{\cal P}_{\cal R}
  (k_{0})} +  \left[ n(k_{0})-1 \right] \ln{\left(\frac{k}{k_{0}}\right)} 
 \nonumber \\
  &+&
   \frac{1}{2} \alpha(k_{0}) \ln^2{\left(\frac{k}{k_{0}}\right)} + ... \,,
\end{eqnarray}
where the spectral index and its running, $\alpha(k) \equiv {\rm d}
\ln n/{\rm d} \ln k $, are evaluated at $k_{0}$ and can be
expressed in terms of the slow-roll parameters. This expansion is
valid provided $\ln{(k/k_{0})}$ is small for the relevant $k$
values. This is the case for cosmological observations, but not for
the wide range of scales probed by PBH constraints.

 In fact only very specific, and contrived, inflaton potentials
produce a constant spectral index~\cite{Vallinotto:2003vf}.  It is
possible for the power spectrum to vary sufficiently with scale so that
PBHs can potentially be over-produced. For instance in the
running-mass inflation model~\cite{Stewart:1996ey,Stewart:1997wg},
the power spectrum is strongly scale-dependent and PBH constraints
exclude otherwise viable regions of parameter
space~\cite{Leach:2000ea,Kohri:2007qn}. More generally, Peiris and
Easther~\cite{Peiris:2008be} recently found, using slow roll
reconstruction, that inflation models which are consistent with
cosmological data can  over-produce PBHs.

Motivated by this, we compile and update the
constraints on the abundance of PBHs (Sec.~\ref{abund}). We then
translate the abundance constraints into detailed generalised
constraints on the power spectrum of the curvature perturbations
(Sec.~\ref{const}), taking into account the evolution of the density
perturbations prior to horizon entry.

\section{PBH abundance constraints}

\label{abund}

The PBH constraints can, broadly, be split into two classes: those
that arise from their present day gravitational consequences and those
that arise from the products of their evaporation. In both cases, in
order to constrain the primordial density or curvature perturbation,
we need to translate the constraints into limits on the initial
PBH mass fraction.

Throughout we will assume that the PBHs form at a single epoch and
their mass is a fixed fraction, $f_M$, of the horizon mass $M_{\rm
PBH}= f_{M} M_{\rm H}$, where $f_{M} \approx (1/3)^{3/2}$~\cite{Carr:1985wss}. 
A scale invariant power spectrum
produces an extended PBH mass function ${\rm d}n_{\rm PBH}/ {\rm d}M_{\rm PBH} \propto M_{\rm
PBH}^{-5/2}$~\cite{Carr:1975qj,MacGibbon:1991vc}, however (as
discussed in the introduction above) in this case the number density
of PBHs would be completely
negligible~\cite{Carr:1993aq,Kim:1999iv}. For scale-dependent power
spectra which produce an interesting PBH abundance it can be assumed
that all PBHs form at a single epoch~\cite{Green:1999xm}.  As a
consequence of near critical phenomena in gravitational
collapse~\cite{Choptuik:1992jv,Gundlach:2002sx,Gundlach:2007gc} the
PBH mass may, however, depend on the size of the fluctuation from
which it forms~\cite{Niemeyer:1997mt,Niemeyer:1999ak,Musco:2004ak} in
which case the mass function has finite width.  Most of the
constraints that we discuss below effectively apply to the mass
function integrated over a range of masses. The range of applicability
is usually significantly larger than the width of the mass function
produced by critical collapse, so in the absence of a concrete
prediction or model for the primordial power spectrum in most cases it
is reasonable to approximate the mass function as a delta-function.
The constraints from cosmic-rays and gamma-rays produced by recently
evaporating PBHs are an exception to this. These constraints depend
significantly on the PBH mass function and therefore need to be
calculated on a case by case
basis~\cite{Kribs:1999bs,Bugaev:2000bz,Bugaev:2002yt,Barrau:2002ru,Bugaev:2008gw}.
We therefore do not include these constraints in our calculation of
generalised constraints on the curvature perturbation power spectrum.

Taking into account the cosmological expansion, the initial PBH mass fraction, $\beta(M_{\rm PBH})$, is related to the present day PBH density, $\Omega_{\rm PBH}^{0}$, by
\begin{equation}
\beta(M_{\rm PBH}) \equiv \frac{\rho_{\rm PBH}^{\rm i}}{\rho_{\rm crit}^{\rm i}}
=\frac{\rho_{\rm PBH}^{\rm eq}}{\rho_{\rm crit}^{\rm eq}}\left(\frac{a_{\rm i}}{a_{\rm eq}}\right)
\approx \Omega_{\rm PBH}^{0} \left(\frac{a_{\rm i}}{a_{\rm eq}}\right)\,,
\end{equation}
where $a$ is the scale factor, `eq' refers to matter-radiation
equality and $\rho_{\rm crit}$ is the critical energy density. Using
the constancy of the entropy, $(s=g_{*s}a^3T^3)$, where $g_{\star s}$
refers to the number of entropy degrees of freedom, we relate the
scale factor to the temperature of the Universe and using the
radiation density, $\rho=\frac{\pi^2}{30}g_\star T^4$, and horizon mass, $M_{\rm
H}=\frac{4 \pi}{3} \rho H^{-3}$, we obtain
\begin{equation}
\beta(M_{\rm PBH}) = \Omega_{\rm PBH}^{0} \left( \frac{g_{\star}^{\rm eq}}{g_{\star}^{\rm i}} 
   \right)^{1/12}  \left( \frac{M_{\rm H}}{M_{\rm H}^{\rm eq}} \right)^{1/2} \,,
\end{equation}
where $g_{\star}$ is the total number of effectively massless degrees of freedom and we have taken
$g_{\star s} \approx g_{\star}$. The horizon mass at matter-radiation
equality is given by (c.f. Ref.~\cite{Green:2004wb})
\begin{equation}
M_{\rm H}^{\rm eq} = \frac{4 \pi}{3} \rho_{\rm eq} H_{\rm eq}^{-3}
          = \frac{8 \pi}{3} \frac{\rho_{\rm rad}^{0}}{a_{\rm eq} k_{\rm eq}^3} \,.
\end{equation} 
Inserting numerical
values, $g_{\star}^{\rm i} \approx 100$, $g_{\star}^{\rm eq}
\approx 3$, $\Omega_{\rm rad}^{0} h^2 = 4.17 \times 10^{-5}$,
$\rho_{\rm crit}  = 1.88 \times 10^{-29}  h^2 \, {\rm g \, cm}^{-3}$,
$k_{\rm eq} = 0.07 \, \Omega_{\rm m}^{0} h^2 \, {\rm Mpc}^{-1}$, $a_{\rm
eq}^{-1} = 24 000 \, \Omega_{\rm m}^{0} h^2$ and $\Omega_{\rm m}^{0} h^2 =
0.1326 \pm 0.0063$~\cite{Dunkley:2008ie} gives  $M_{\rm H}^{\rm eq}= 1.3 \times
10^{49} (\Omega_{\rm m} h^2)^{-2} \, {\rm g}$
so that
\begin{equation}
\label{betaomega}
\beta(M_{\rm {PBH}}) = 6.4 \times 10^{-19} \, \Omega_{\rm PBH}^{0} 
   \left( \frac{M_{\rm PBH}}{f_{M} 5\times10^{14} \, {\rm g}} \right)^{1/2} \,.
\end{equation}

As first shown by Hawking~\cite{Hawking:1974rv} a black hole with mass
$M_{\rm PBH}$ emits thermal radiation with temperature
\begin{equation}
T_{\rm BH} = \frac{1}{8 \pi G M_{\rm PBH}}  \approx 1.06  
 \left( \frac{10^{13} \, {\rm g}}{M_{\rm PBH}} \right) \, {\rm GeV}\,.
\end{equation}
The current understanding of PBH evaporation~\cite{MacGibbon:1990zk} is that
PBHs directly emit all particles which appear elementary at the energy
scale of the PBH and have rest mass less than the black hole
temperature. Thus if the black hole temperature exceeds the QCD
confinement scale, quark and gluon jets are emitted directly. The
quark and gluon jets then fragment and decay producing astrophysically
stable particles: photons, neutrinos, electrons, protons and their
anti-particles.  Taking into account the number of emitted species the
mass loss rate can be written as~\cite{MacGibbon:1991tj}
\begin{equation}
\frac{{\rm d} M_{\rm PBH}}{{\rm d} t} 
  = - 5.34 \times 10^{25} \phi(M_{\rm PBH}) M_{\rm PBH}^{-2} \, 
     {\rm g \, s}^{-1} \,,
\end{equation}
where $\phi(M_{\rm PBH})$ takes into account the number of directly
emitted species ($\phi(M_{\rm PBH}) = 0.267 g_{0} + 0.147 g_{1/2} + 0.06
g_{1}+ 0.02 g_{3/2} +0.007 g_{2}$ where $g_{\rm s}$ is the number of
degrees of freedom with spin $s$) and is normalized to one for PBHs
with mass $M_{\rm PBH} \gg 10^{17} \, {\rm g}$ which can only emit
photons and neutrinos.  For lighter PBHs $\phi(5 \times 10^{14} \,
{\rm g} < M_{\rm PBH} < 10^{17} \, {\rm g})=1.569$.
The PBH lifetime is then given by~\cite{MacGibbon:1991tj}
\begin{equation}
\tau \approx 6.24 \times 10^{-27} M_{\rm PBH}^3 \phi(M_{\rm PBH})^{-1} 
     \, {\rm s} \,.
\end{equation}

From the WMAP 5 year data~\cite{Dunkley:2008ie} the present age of the
Universe is $t_0=13.69 \pm 0.13$ Gyr.  The initial mass of a PBHs
which is evaporating today is therefore $ M_{\rm PBH}\approx
5\times10^{14}$g~\cite{MacGibbon:2007yq}, while less massive PBHs will
have evaporated by the present day.

We will now compile, and where relevant update, the PBH abundance
constraints.  We divide the constraints into two classes: those, for
PBHs with $M_{\rm PBH} > 5 \times 10^{14}$g, arising from their
gravitational consequences (Sec.~\ref{gravconst}) and those for
$M_{\rm PBH} < 5 \times 10^{14}$g arising from their evaporation
(Sec.~\ref{evapconst}).

\subsection{Gravitational constraints}

\label{gravconst}

\subsubsection{Present day density}

The present day density of PBHs with $M_{\rm PBH} > 5 \times 10^{14}
\, {\rm g}$ which haven't evaporated by today must be less than the upper limit on the present day cold dark matter (CDM) density. Using the 5 year WMAP measurements~\cite{Dunkley:2008ie},
$\Omega_{\rm CDM}^{0} h^2 = 0.1099 \pm 0.0062$,
$h=0.719^{+0.026}_{-0.027}$, gives
\begin{equation}
\Omega_{\rm PBH}^{0} <  0.25   \,,
\end{equation}
which, using Eq.(\ref{betaomega}), leads to
\begin{eqnarray}
\beta(M_{\rm PBH}) &<& 1.6 \times 10^{-19} 
   \left( \frac{M_{\rm PBH}}{f_{M}\, 5\times10^{14} \, {\rm g}} \right)^{1/2} 
    \nonumber \\
 && \,\, {\rm for} \,\,  M_{\rm PBH} > 5\times10^{14} \, {\rm g} \,.
\end{eqnarray}

\subsubsection{Lensing of cosmological sources}

If there is a cosmologically significant density of compact objects
then the probability that a distant point source is lensed is
high~\cite{pressgunn:1973ApJ}. The limits as given below have been
calculated assuming an Einstein de Sitter Universe, $\Omega_{\rm
m}=1$, and a uniform density of compact objects. The recalculation
of the constraints for a $\Lambda$ dominated Universe would be
non-trivial. The constraints would, however, be tighter (due to the
increased path length and the larger optical depth to a given
red-shift)~\cite{Dalcanton:1994ApJ}, and the constraints given below
are therefore conservative and valid to within a factor of order
unity.

\subsubsection*{Gamma-ray bursts}

Light compact objects can femtolens gamma-ray bursts (GRBs), producing a
characteristic interference pattern~\cite{Gould:1992td}. A null search
using BATSE data leads to a
constraint~\cite{Marani:1998sh}
\begin{equation}
\Omega_c < 0.2 \,\,\, {\rm for} \,\, 10^{-16} M_{\odot} < M_{\rm PBH} < 10^{-13} M_{\odot} \,,
\end{equation}
where $\Omega_c$ is the density of compact objects, assuming a mean GRB red-shift of one.

\subsubsection*{Quasars}
Compact objects with mass $10^{-3} M_{\odot} < M_{\rm PBH} < 300 M_{\odot}$ can
microlens quasars, amplifying the continuum emission without
significantly changing the line emission~\cite{Canizares:1982ApJ}.  Limits on an
increase in the number of small equivalent width quasars with
red-shift lead to the constraint~\cite{Dalcanton:1994ApJ}:
\begin{equation}
\Omega_c < 0.2 \,\,\,\, {\rm for} \,\, 0.001 M_{\odot} < M_{\rm PBH} < 60 M_{\odot} \,,
\end{equation}
assuming $\Omega_{\rm tot}= \Omega_c$.

\subsubsection*{Radio sources}

Massive compact objects, $10^{6} M_{\odot} < M_{\rm PBH} < 10^{8} M_{\odot}$,
can millilens radios sources producing multiple sources with
milliarcsec separation~\cite{Kassiola:1990jj}. A null search using 300 compact
radio sources places a constraint~\cite{Wilkinson:2001vv}
\begin{equation}
\Omega_c < 0.013  \,\,\,{\rm for}\,\,  10^{6} M_{\odot} < M_{\rm PBH} < 10^{8} M_{\odot} \,.
\end{equation}

\subsubsection{Halo fraction constraints}

There are also constraints from the gravitational consequences of PBHs
within the Milky Way halo. They are typically expressed in terms of
the fraction of the mass of the Milky Way halo in compact objects,
$f_{\rm h} = M_{\rm PBH}^{\rm MW}/M_{\rm tot}^{\rm MW}$. They require
some modeling of the Milky Way halo (typically the density and/or
velocity distribution of the halo objects). Consequently there is
a factor of a few uncertainty in the precise values of the
constraints.

Assuming that PBHs make up the same fraction of the halo dark matter
as they do of the cosmological cold dark matter, and ignoring the
uncertainties in the CDM density (since this is negligible compared
with the uncertainties in halo fraction limit calculations), we can
relate the halo fraction to the PBH cosmological density:
\begin{equation}
f_{h} \equiv \frac{ M_{\rm PBH}^{\rm MW}}{M_{\rm CDM}^{\rm MW} }
   \approx \frac{\rho_{\rm PBH}^{0}}{\rho_{\rm CDM}^{0}}
   = \frac{\Omega_{\rm PBH}^{0} h^2}{\Omega_{\rm CDM}^{0} h^2}  \approx 5 \Omega_{\rm PBH}^{0}
\,.
\end{equation}

\subsubsection*{Microlensing}

Solar and planetary mass compact objects in the Milky Way halo can
microlens stars in the Magellanic Clouds, causing temporary one-off
brightening of the microlensed star~\cite{Paczynski:1985jf}.  The
relationship between the observed optical depth, $\tau$, (the
probability that a given star is amplified by more than a factor of
1.34) and the fraction of the halo in MACHOs depends on the
distribution of MACHOS in the MW halo. For the `standard' halo model
used by the microlensing community (a spherical cored isothermal
sphere) $\tau \approx 5 \times 10^{-7} f_{h}$~\cite{Griest:1990vu,Alcock:2000ph}, with the derived value of limits on $f_{h}$ varying by factors of order unity for other halo
models.

The EROS collaboration find a $95\%$ upper confidence limit $\tau <
0.36 \times 10^{-7}$ which they translate into limits on the halo
fraction~\cite{Tisserand:2006zx}:
\begin{equation}
f_{h} < 0.04 \,\,\,\,\,\, {\rm for} \,\,\,\, 10^{-3} M_{\odot} < M_{\rm PBH}< 10^{-1} M_{\odot} \,, \\
\end{equation}
or
\begin{equation}
f_{h}   < 0.1 \,\,\,\,\,\, {\rm for} \,\,\,\, 10^{-6} M_{\odot} < M_{\rm PBH} < M_{\odot} \,.
\end{equation}
Combined EROS and MACHO collaboration limits on short duration
events constrain the abundance of light MACHOs~\cite{Alcock:1998fx} 
\begin{equation}
f_{h}  <  0.25 \,\,\,\,\,\, {\rm for} \,\,\,\, 10^{-7} M_{\odot} < M_{\rm PBH} < 10^{-3} M_{\odot} \,,
\end{equation}
while a dedicated search by the MACHO collaboration
for long ($> 150$ days) duration events leads to 
limits on more massive MACHOs~\cite{Allsman:2000kg}:
\begin{equation}
f_{h} < 1.0  \,\,\,\,\,\, {\rm for} \,\,\,\,  0.3 M_{\odot} < M_{\rm PBH} < 30 M_{\odot} \,,
\end{equation}
or
\begin{equation}
f_{h} < 0.4   \,\,\,\,\,\, {\rm for} \,\,\,\,     M_{\rm PBH} < 10 M_{\odot} \,.
\end{equation}

Combined, these limits give
\begin{eqnarray}
f_{h}  & <& 0.25 \,\,\,\, {\rm for} \,\, 10^{-7} M_{\odot} < M_{\rm PBH} < 10 ^{-6} M_{\odot} \,, \\
f_{h}  & <& 0.1 \,\,\,\,\, {\rm for} \,\,\, 10^{-6} M_{\odot} < M_{\rm PBH} < M_{\odot} \,, \\
f_{h}  & <& 0.4 \,\,\,\,\, {\rm for} \,\,\,  M_{\odot} < M_{\rm PBH} < 10 M_{\odot} \,.
\end{eqnarray}

\subsubsection*{Wide binary disruption}

More massive compact objects would affect the orbital parameters of
wide binaries~\cite{Bahcall:1984ai,Weinberg:1987ApJ}. Comparison of
the separations of observed halo binaries with simulations of
encounters between compact objects and wide binaries lead to a
constraint~\cite{Yoo:2003fr}
\begin{equation}
f_{h} < 0.2 \,\,\,\,\,\, {\rm for} \,\,\,\, 10^{3} M_{\odot} < M_{\rm PBH} < 10^{8} M_{\odot} \,.
\end{equation}
See however Ref.~\cite{Quinn:2009zg} for a recent re-examination of
this constraint which leads to a somewhat weaker limit.

\subsubsection*{Disk heating}

Massive halo objects traversing the Galactic disk will heat the disk,
increasing the velocity dispersion of the disk
stars~\cite{1985ApJ...299..633L}. This leads to a limit, from the observed
stellar velocity dispersions, on the halo fraction in massive
objects~\cite{Carr:1997cn}
\begin{eqnarray}
f_{h} &<& \frac{M_{\rm disk, lim}}{M_{\rm PBH}}\,, \nonumber \\
M_{\rm disk, lim} &=& 3 \times 10^{6} \left(\frac{\rho_{h}}{0.01 M_{\odot} \rm{pc}^{-3}}
\right)^{-1}     \nonumber \\
&\times& \left( \frac{\sigma_{\rm obs}}{60 \, {\rm km \, s}^{-1}}
 \right)^2 \left( \frac{t_{\rm s}}{10^{10} \, {\rm yr}} \right)^{-1} M_{\odot}\,,
\end{eqnarray}
where $\rho_{h}$ is the local halo density and $\sigma_{\rm obs}$ and
$t_{\rm s}$ are the velocity dispersion and age of the halo stars, respectively.

\subsection{Evaporation constraints}

\label{evapconst}

\subsubsection{Diffuse gamma-ray background}

PBHs with masses in the range $ 2 \times 10^{13} < M_{\rm PBH}< 5 \times
10^{14} {\rm g}$ evaporate between $z \approx 700$ and the present day
and can contribute to the diffuse gamma-ray
background~\cite{Kim:1999iv,Carr:1976ApJ,Page:1976wx,MacGibbon:1991vc,Halzen:1991uw,Kribs:1999bs}. As discussed above, these constraints depend significantly on the PBH mass
function and hence we will not consider them further.

\subsubsection{Cosmic-rays}

The abundance of PBHs evaporating around the present day can also be
constrained by limits on the abundance of cosmic-rays (in particular
positrons and antiprotons)~\cite{MacGibbon:1991vc,Yoshimura:2001}. The
constraints from anti-protons have been calculated for several mass
functions and are essentially equivalent to those from the diffuse
gamma-ray background~\cite{Barrau:2002mc,Barrau:2002ru}.

\subsubsection{Neutrinos}

Neutrinos produced by PBH evaporation contribute to the
diffuse neutrino background.  The neutrino spectrum, and hence the
resulting PBH abundance constraints, depend strongly on the PBH mass
function, but the constraints are typically weaker than those from the
diffuse gamma-ray background~\cite{Bugaev:2000bz,Bugaev:2002yt}.

\subsubsection{Hadron injection}

PBHs with mass $M_{\rm PBH} < 10^{10} \, {\rm g}$ have a lifetime
$\tau < 10^{3} \, {\rm s}$ and evaporate before the end of
nucleosynthesis, and can therefore affect the light element
abundances~\cite{Vainer:1977PAZh,Miyama:1978mp,Zeldovich:1977SvAL,Vainer:1978SvAL}.
The constraints from hadron injection have been re-evaluated 
(see Ref.~\cite{Kohri:1999ex}), taking into account the emission of 
fundamental particles and using more up to date measurements of the 
Deuterium and ${}^4$He abundances ($D/H\leq 4.0 \times 10^{-5}, Y_{p} \leq 0.252$
respectively):
\begin{eqnarray}
\beta(M_{\rm PBH})&<&10^{-20} \,\,\,\, {\rm for} \,\,
         10^{8} {\rm g} < M_{\rm PBH}<10^{10} {\rm g}  \,, \\
\beta(M_{\rm PBH})&<&10^{-22} \,\, \nonumber \\  && {\rm for} \,\, 
        10^{10} {\rm g} < M_{\rm PBH}<3\times 10^{10} {\rm g}.
\end{eqnarray}

\subsubsection{Photodissociation of deuterium}

The photons produced by PBHs which evaporate between the end of
nucleosynthesis and recombination can photodissociate
deuterium~\cite{Lindley:1980MNRAS}. The resulting constraints on the
PBH abundance have been updated, in the context of braneworld
cosmology in Ref.~\cite{Clancy:2003zd}. Adapting that calculation to
the standard cosmology we find:
\begin{eqnarray}
\beta(M_{\rm PBH}) &<& 3 \times 10^{-22} \left(\frac{M_{\rm PBH}}
   {f_M 10^{10}g}\right)^{1/2} \nonumber \\          && 
    \,\, {\rm for} \,\,    10^{10} {\rm g} < M_{\rm PBH}< 10^{13} {\rm g}.
\end{eqnarray}

\subsubsection{CMB distortion}

Photons emitted by PBHs which evaporate between $z\sim 10^{6}$ and
recombination at $z\sim 10^{3}$ can produce distortions in the cosmic
microwave background radiation~\cite{Naselskii:1978SvAL}. Using the
COBE/FIRAS limits on spectral distortions of the CMB from a black body
spectrum~\cite{Mather:1993ij}, Ref.~\cite{Tashiro:2008sf} finds
\begin{equation}
\beta(M_{\rm PBH}) < 10^{-21} \,\,\,\, {\rm for} \,\,\, 
     10^{11} \, {\rm g} < M_{\rm PBH} < 10^{13} \, {\rm g} \,.
\end{equation}

\subsubsection{(Quasi-)stable massive particles}

In extensions of the standard model there are generically stable or
long lived massive (${\cal O}(100 \, {\rm GeV})$) particles. Light
PBHs with mass $M_{\rm PBH} \lesssim 10^{11} \, {\rm g}$ can emit
these particles and their abundance is hence limited by the present
day abundance of stable massive particles~\cite{Green:1999yh} and the
decay of long-lived
particles~\cite{Lemoine:2000sq,Khlopov:2004tn}\footnote{More massive
PBHs can also emit these particles in the late stages of their
evaporation, when their mass drops below $\sim 10^{9} \, {\rm g}$.
However the resulting constraints are substantially weaker than those
from hadron injection during nucleosynthesis.}.

Gravitinos in supergravity theories and moduli in string theories are
generically quasi-stable and decay after big bang nucleosynthesis,
potentially altering the light element abundances. 
The effect of their decay on the products of big bang nucleosynthesis leads 
to a constraint on the initial PBH fraction~\cite{Lemoine:2000sq}:
\begin{eqnarray}
\beta(M_{\rm PBH})&<& 5\times10^{-19} \left(\frac{g_{\star}^{\rm i}}{200} \right)^{1/4}
     \left(\frac{\alpha}{3} \right)
    \left( \frac{x_{\phi}}{6 \times 10^{-3}}\right)    \nonumber
      ^{-1}\\          \nonumber
     &\times& \left( \frac{f_{M} M_{\rm PBH}}{10^{9} \, {\rm g} }\right)^{-1/2}
      \left( \frac{\bar{Y_{\phi}}}{10^{-14}} \right)       \\
  && \,\, {\rm for} \,\,      M_{\rm PBH} < 10^{9} \, {\rm g},
\end{eqnarray}
where $x_{\phi}$ is the fraction of the luminosity going into
quasi-stable massive particles, $g_{\star}^{\rm i}$ is the initial
number of degrees of freedom (taking into account supersymmetric
particles), $\alpha$ is the mean energy of the particles emitted in
units of the PBH temperature and $\bar{Y_{\phi}}$ is the limit
on the quasi-stable massive particle number density to
entropy density ratio.

In supersymmetric models, in order to avoid the decay of the proton, there
is often a conserved quantum number R-parity, which renders the
Lightest Supersymmetric Particle (LSP) stable and the present day density of
such stable particles produced via PBH evaporation must not exceed the
upper limit on the present day CDM density~\cite{Green:1999yh}.  This
leads to a constraint on the initial PBH fraction
(c.f.~Ref.~\cite{Lemoine:2000sq}):
\begin{eqnarray}
\beta(M_{\rm PBH}) &<& 6 \times 10^{-19} h^2 \left(\frac{g_{\star}^{\rm i}}{200} 
   \right)^{1/4}
     \left(\frac{\alpha}{3} \right) \left( \frac{x_{\rm LSP}}{0.6} \right)^{-1} 
  \nonumber  \\
     &\times& \left( \frac{f_{M} M_{\rm PBH}}{10^{11} \, {\rm g} }\right)^{-1/2}
      \left( \frac{m_{\rm LSP}}{100 \, {\rm GeV}} \right)^{-1}      \nonumber  \\
       && \,\, {\rm for} \,\,  M_{\rm PBH} < 10^{11} \, {\rm g}\left(\frac{100\,{\rm GeV}}
          {m_{\rm LSP}}\right) ,
\end{eqnarray}
where  $m_{\rm LSP}$ is the
mass of the LSP and $x_{\rm LSP}$ is the fraction of the
luminosity carried away by the LSP.

These constraints depend on the (uncertain) details of physics beyond
the standard model, and we therefore summarise them conservatively as
\begin{eqnarray}
\beta(M_{\rm PBH})  &\lesssim& 10^{-18}
\left( \frac{f_{M} M_{\rm PBH}}{10^{11}{\rm g}} \right)^{-1/2}		\nonumber \\
      &&   {\rm for} \,\,  M_{\rm PBH} < 10^{11} \, {\rm g} \,.
\end{eqnarray}

\subsubsection{Present day relic density}

It has been argued that black hole evaporation could leave a stable
Planck mass relic~\cite{Bowick:1988xh,Coleman:1991sj,Macgibbon:1987my}, in which
case the present day density of relics must not exceed the upper limit
on the CDM density
\begin{equation}
\Omega_{\rm rel}^{0} < 0.25 \,.
\end{equation}
Writing the relic mass as $M_{\rm rel} = f_{\rm rel} M_{\rm Pl}$ this 
gives a tentative constraint
\begin{eqnarray}
 \beta(M_{\rm PBH}) &<& 4 \frac{1}{f_{M}^{1/2} f_{\rm rel}} 
   \left( \frac{M_{\rm PBH}}{ 5\times10^{14} \, {\rm g}} \right)^{3/2}  \
   \nonumber \\  && \,\, {\rm for} \,\,  M_{\rm PBH} < 5\times10^{14}
        \, {\rm g}  \,.
\end{eqnarray}

\vspace{1.0cm}

The constraints are summarised in table~\ref{consttable} and are
displayed in Fig.~1.  As can be seen from Fig.~1, the constraints probe 
a very large range of scales and in some cases several constraints overlap 
across particular mass ranges.  The solid line indicates the strongest 
constraints for each mass scale and we consider only these when constraining 
the primordial power spectrum in Sec.~\ref{const}.

\begin{table*}[t]
\begin{center}
\caption{Summary of constraints on the initial PBH abundance,
 $\beta(M_{\rm PBH})$.}
\label{consttable}
\medskip
\begin{tabular}{|c|c|c|}
\hline
description &  mass range & constraint on $\beta(M_{\rm PBH})$ \\ \hline

\multicolumn{3}{|c|}{Gravitational constraints} \\ \hline
present day PBH density & $M_{\rm PBH} > 5 \times 10^{14} \, {\rm g}$& 
$ < 2 \times 10^{-19}\left( \frac{M_{\rm PBH}}{f_{M} 5 \times 10^{14} \, 
{\rm g}} \right)^{1/2}$ \\ \hline

GRB femtolensing & $ 10^{-16} M_{\odot} < M_{\rm PBH} < 10^{-13} M_{\odot} $  & 
 $ < 1 \times 10^{-19} \left( \frac{M_{\rm PBH}}{f_{M} 5 \times 10^{14} \, 
{\rm g}} \right)^{1/2}$ \\ \hline
Quasar microlensing &$0.001 M_{\odot} < M_{\rm PBH} < 60 M_{\odot}$ & 
$ < 1 \times 10^{-19} 
   \left( \frac{M_{\rm PBH}}{f_{M} 5 \times 10^{14} \, {\rm g}} 
\right)^{1/2}$ \\ \hline
Radio source microlensing &  $ 10^{6} M_{\odot} < M_{\rm PBH} < 10^{8} M_{\odot} $  
&   $ < 6 \times 10^{-20} 
   \left( \frac{M_{\rm PBH}}{f_{M} 5 \times 10^{14} \, {\rm g}} \right)^{1/2}$
  \\ \hline
\multicolumn{3}{|c|}{Halo density \footnote{These constraints
depend on the PBH distribution within the Milky Way halo and hence have a 
factor of order a few 
uncertainty.}} \\ \hline
LMC Microlensing 
  & $10^{-7} M_{\odot} < M_{\rm PBH} < 10 ^{-6} M_{\odot}$    & $< 3 \times
 10^{-20}  \left( \frac{M_{\rm PBH}}{f_{M} 5 \times 10^{14} \, {\rm g}} 
\right)^{1/2} $  \\
 & $ 10^{-6} M_{\odot} < M_{\rm PBH} < M_{\odot}$ &   $<1 \times 10^{-20}
  \left( \frac{M_{\rm PBH}}{f_{M} 5 \times 10^{14} \, {\rm g}} 
\right)^{1/2} $  \\
 & $ M_{\odot} < M_{\rm PBH} < 10 M_{\odot}$ &  $< 5 \times 10^{-20}
  \left( \frac{M_{\rm PBH}}{f_{M} 5 \times 10^{14} \, {\rm g}}
 \right)^{1/2}$   \\ \hline
Wide binary disruption & $ 10^{3} M_{\odot} < M_{\rm PBH} < 10^{8} M_{\odot}$
 & $<3 \times 10^{-20}  \left( 
\frac{M_{\rm PBH}}{f_{M} 5 \times 10^{14} \, {\rm g}}
 \right)^{1/2}$ \\ \hline
Disk heating 
& $M_{\rm PBH}> 3 \times 10^{6} M_{\odot}$ &  
$<2 \times 10^{6} \frac{1}{f_{M}^{1/2}} \left( 
\frac{M_{\rm PBH}}{ 5 \times 10^{14} \, {\rm g}}
 \right)^{-1/2}$
\\ \hline 
\multicolumn{3}{|c|}{Evaporation} \\ \hline
diffuse gamma-ray background & $ 2 \times 10^{13} \, {\rm g}
< M_{\rm PBH}< 5 \times 
  10^{14} \, {\rm g}$ & {\em depends on PBH mass function} \\ \hline
cosmic-rays & {\rm similar to DGRB} & {\em depends on PBH mass 
function} \\ \hline
neutrinos & {\rm similar to DGRB} & {\em depends on PBH mass 
function}\\ \hline
hadron injection & $ 10^{8} \, {\rm g} < M_{\rm PBH} < 10^{10} \, {\rm g} $
  & $< 10^{-20}$ \\
   & $10^{10} \, {\rm g} <M_{\rm PBH} < 3 \times 10^{10} \, {\rm g}$ & $
 <10^{-22}$\\ \hline
photodissociation of deuterium & $10^{10} \, {\rm g} <M_{PBH}<
10^{13}\, {\rm g} $ & $ < 3 \times
 10^{-22} \left( \frac{M_{PBH}}{f_M 10^{10}g} \right)^{1/2}$ \\ \hline
CMB distortion & $10^{11} \, {\rm g} < M_{\rm PBH} < 10^{13} \, {\rm g}$ & 
$< 10^{-21}$ \\ \hline
(Quasi-)stable massive particles \footnote{Conservative summary, depends on 
physics beyond the standard model of particle physics.}  & 
$M_{\rm PBH} < 10^{11} \, {\rm g} $
& $< \sim 10^{-18}
\left( \frac{f_{M} M_{\rm PBH}}{10^{11} \, {\rm g}} \right)^{-1/2}$\\ \hline
present day relic density \footnote{Only applies if evaporation leaves stable
relic.} &
$M_{\rm PBH} < 5 \times 10^{14} \, {\rm g}$& $ < 4 \frac{1}{f_{M}^{1/2
} f_{\rm rel}} \left( \frac{M_{\rm PBH}}{ 5 \times 10^{14} \, {\rm g}} \right)^{3/2}$  
  \\ \hline
\hline
\end{tabular}
\end{center}
\end{table*}

\section{Constraints on the curvature perturbation power spectrum}

\label{const}

We focus in the following on the standard case of PBH formation, which
applies to scales which have left the horizon at the end of
inflation. It has recently been
shown~\cite{Lyth:2005ze,Zaballa:2006kh} that PBHs can also form on
scales which never leave the horizon during inflation, and therefore
never become classical. We also only consider gaussian perturbations
and a trivial initial radial density profile, and refer to
Ref.~\cite{Hidalgo:2007vk} for the effects of non-gaussian
perturbations and to Refs.~\cite{Shibata:1999zs,Hidalgo:2008mv} for
estimates on the effect of deviations from a trivial initial density
profile.

A region will collapse to form a PBH if the smoothed density contrast,
in the comoving gauge, at horizon crossing ($R= (a
H)^{-1}$), $\delta_{\rm hor}(R)$, satisfies the
condition $\delta_{\rm c} \leq\delta_{\rm hor}(R)\leq 1$~\cite{Carr:1985wss},
where $\delta_{\rm c} \sim 1/3$. The mass of the PBH formed is
approximately equal to the horizon mass at horizon entry, $M_{\rm
PBH}= f_{M} M_{\rm H}$, which is related to the smoothing scale,
$R$, by~\cite{Green:2004wb}
\begin{equation}
M_{\rm H} = M_{\rm H}^{\rm eq} (k_{\rm eq} R)^2 \left( \frac{g_{\star, {\rm eq}}}{
   g_{\star}} \right)^{1/3} \,,
\end{equation}
where  $M_{\rm H}^{\rm eq}= 1.3 \times 10^{49} (\Omega_{\rm m} h^2)^{-2} \, {\rm g}$ is the horizon mass at
matter-radiation equality.

Taking the initial perturbations to be Gaussian, the probability
distribution of the smoothed density contrast, $P(\delta_{\rm
hor}(R))$, is given by (e.g. Ref.~\cite{Liddle:2000cg})
\begin{equation}
P(\delta_{\rm hor}(R))=\frac{1}{\sqrt{2\pi} \sigma_{\rm hor}(R)} \exp{ \left(
    - \frac{\delta_{\rm hor}^2(R)}{2 \sigma_{\rm hor}^2(R)} \right)} \,,
\end{equation}
where $\sigma(R)$ is the mass variance
\begin{equation}
\label{variance}
\sigma^2(R)=\int_{0}^{\infty} W^2(kR)\mathcal{P}_{\delta}(k, t)\frac{\,dk}{k},
\end{equation}
and $W(kR)$ is the Fourier transform of
the window function used to smooth the density contrast. We assume a Gaussian
window function for which $W(kR)= \exp{(-k^2 R^2/2)}$.

The fraction of the energy density of the Universe contained in
regions dense enough to form PBHs is then given, as in
Press-Schechter theory~\cite{Press:1973iz} by,
\begin{equation}
\beta(M_{\rm PBH})= 2 \frac{M_{\rm PBH}}{M_{H}} 
      \int_{\delta_{\rm c}}^{1} P(\delta_{\rm hor}(R)) \,d\delta_{\rm hor}(R) \,.
	               \label{presssch}
\end{equation}
This leads to a relationship between the PBH
initial mass fraction and the mass variance, 
\begin{eqnarray}	
\beta(M_{\rm PBH})& = & \frac{2 f_{\rm M}}{\sqrt{2\pi}\sigma_{\rm hor}(R)} 
 \nonumber
   \\
  && \times 
\int_{\delta_{\rm c}}^{1} \exp{\left(- \frac{\delta^2_{\rm hor}(R)}
    {2 \sigma_{\rm hor}^2(R)}\right)} 
  \,d\delta_{\rm hor}(R) \,,\nonumber\\
& \approx & f_{M} {\rm erfc}\left(\frac{\delta_{\rm c}}{
   \sqrt{2}\sigma_{\rm hor}(R)}\right) \,. \label{densitypara}
\end{eqnarray}
The constraints on the PBH initial mass fraction can therefore be
translated into constraints on the mass variance by simply inverting
this expression.


\begin{figure}
\begin{center}
\label{allconstraintsplot}
\includegraphics[width=8.5cm]{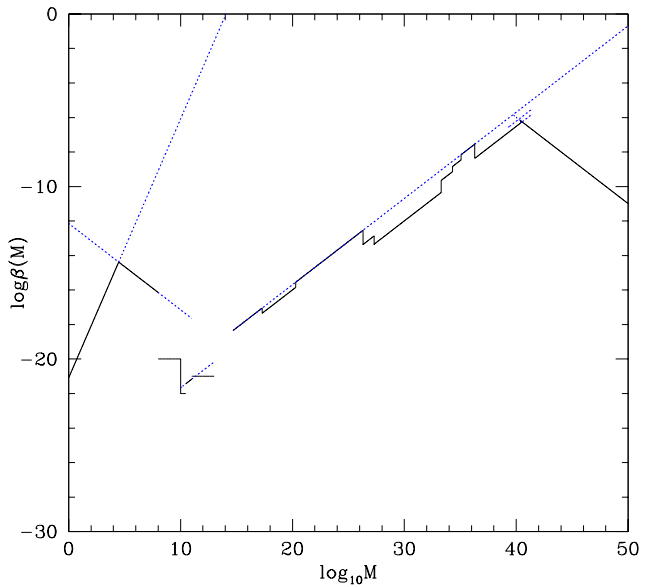}
\caption{The limits on the initial mass fraction of PBHs as a function
of PBH mass (in grams). The solid lines represent the tightest limits
for each mass range and the dotted lines are the weaker limits where
there is an overlap between constraints.  As discussed in Sec.~\ref{abund} we have not considered the diffuse gamma-ray background constraint which applies for $2 \times 10^{13} \, {\rm g}
< M_{\rm PBH}< 5 \times 10^{14} \, {\rm g}$ as it depends significantly on the PBH mass function.}
\end{center}
\end{figure}

In order to calculate the mass variance we need the density contrast
in the comoving, or total matter, gauge as a function of time and
scale (c.f.~Ref.~\cite{Bringmann:2001yp,Bugaev:2008gw}). To calculate
this we take the expressions for the evolution of perturbations in the
conformal Newtonian gauge, valid on both sub- and super-horizon
scales, and carry out a gauge transformation to the total matter gauge
(for further details see Appendix~\ref{gauge}). We find
\begin{equation}
\delta(k,t)=-\frac{4}{\sqrt{3}} \left(\frac{k}{aH}\right) 
     j_{1}(k/\sqrt{3}aH) \mathcal{R} \,,
\end{equation}
where $j_{1}$ is a spherical Bessel function and $\cal{R}$ is the
primordial curvature perturbation. Hence
the power spectrum of the density contrast is given by:
\begin{equation}
\label{powerallscales}
\mathcal{P}_{\delta}(k,t)=\frac{16}{3} \left(\frac{k}{aH}\right)^2 
    j_1^2(k/\sqrt{3} aH)
       \mathcal{P}_{\mathcal{R}}(k) \,.		
\end{equation}
Substituting this into Eq.~(\ref{variance}),
and setting $R=(a H)^{-1}$, gives
\begin{eqnarray}
\label{variancefinal}
\sigma_{\rm hor}^2(R) &=& \frac{16}{3} \int_{0}^{\infty}
     \left( kR \right)^2  j_{1}^2(k R/\sqrt{3}) \nonumber \\
  && \times
 \exp(- k^2 R^2)
         \mathcal{P}_{\mathcal{R}}(k)  \frac{{\rm d} k}{k}\,.
\end{eqnarray}

Since the integral is dominated by scales $k \sim 1/R$ we assume
that, {\em over the scales probed by a specific PBH abundance constraint},
the curvature power spectrum can be written as a power law
\begin{equation}
\label{powerpbh}
{\cal{P}}_{\cal{R}} (k) = {\cal{P}}_{\cal R}(k_0) 
      \left( \frac{k}{k_{0}} \right)^{n(k_{0})-1} \,.
\end{equation}
This assumption is valid for general slow-roll inflation models such
as those considered in
Refs.~\cite{Leach:2000ea,Kohri:2007qn,Peiris:2008be}.  Using
Eqs.~(\ref{densitypara}) and (\ref{variancefinal}) we can translate
the PBH abundance constraints in Sec.~\ref{abund} into constraints on
the amplitude of the curvature perturbation spectrum. 
For each constraint we take the pivot point, $k_{0}$, to correspond to
the scale of interest, $k_{0}=1/R$, and consider a range of values for
$n(k_{0})$ consistent with slow roll inflation, $0.9 < n(k_{0}) <
1.1$. The resulting constraints for $n(k_{0})=1$ are displayed in
Fig.~2.  For $n(k_{0})= 0.9$ and $1.1$ the constraints are weakened or
strenghtened, respectively, at the order of 2 percent. This
indicates that, for slow-roll inflation models, the constraints are
not particularly sensitive to the exact shape of the power spectrum in
the vicinity of the scale of interest. 
The large scale constraints
(small $k$) come from various astrophysical sources such as Milky Way
disk heating, wide binary disruption and a variety of lensing effects.
The small scale constraints generally arise from the consequences of
PBH evaporation, in particular on nucleosynthesis and the CMB.  These
evaporation constraints lead to tighter constraints on the abundance
of PBHs and therefore the primordial power spectrum is more tightly
constrained on these scales.  In general the constraints on the
amplitude of the primordial power spectrum span the range
$\mathcal{P}_{\mathcal{R}}<10^{-2}-10^{-1}$ with some scale
dependence.

\begin{figure}
\begin{center}
\label{powerconstraintsplot}
\includegraphics[width=8.5cm]{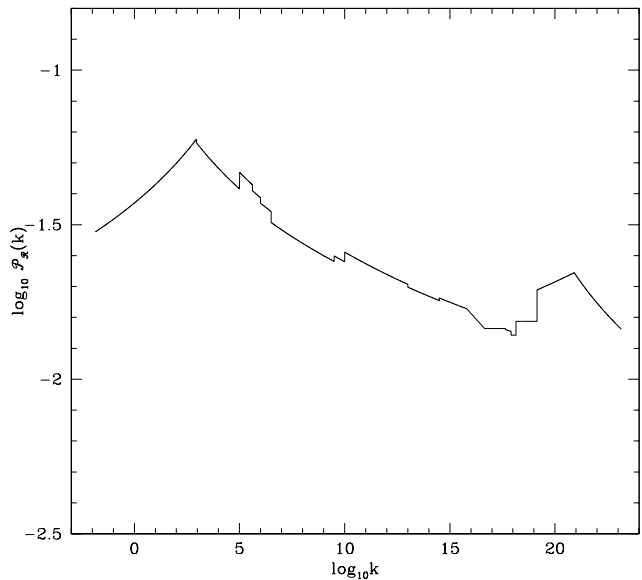}
\caption{Generalised constraints on the amplitude of the power
  spectrum of primordial curvature perturbations as a function of
  comoving wavenumber (in units of ${\rm Mpc}^{-1}$). We have assumed
  that the power spectrum is scale-invariant over the (relatively
  small) range of scale which contribute to a given
  constraint. Deviations from scale-invariance consistent with
  slow-roll inflation lead to small changes in the constraints (see
  text for further details).}
\end{center}
\end{figure}

\section{Summary}

\label{summ}

We have compiled, and where relevant updated, the observational limits
on the initial abundance of Primordial black holes. We then translated
these limits into generalised constraints on the power spectrum of the
primordial curvature perturbation, taking into account the full time
evolution of the density contrast. The constraints on the amplitude of
the power spectrum are typically in the range
$\mathcal{P_{\mathcal{R}}}<10^{-2}-10^{-1}$ with some scale
dependence. This is slightly weaker than the
$\mathcal{P_{\mathcal{R}}}<10^{-3}-10^{-2}$ assumed in
Ref.~\cite{Peiris:2008be}.  These more accurate generalised
constraints could be used to more accurately constrain the parameter
space of slow-roll inflation models
(c.f.~\cite{Carr:1994ar,Green:1997sz,Kim:1996hr,Bringmann:2001yp,Leach:2000ea,Kohri:2007qn,Peiris:2008be}).

\begin{acknowledgments}
  The authors would like to thank Carlos Hidalgo and Jane MacGibbon
  for useful discussions and comments. AJ is supported by the
  University of Nottingham, AMG is supported by STFC.
\end{acknowledgments}

\appendix
\section{Density contrast calculation}
\label{gauge}

The general, scalar, perturbed metric can be written as 
(c.f.~\cite{Liddle:2000cg}):
\begin{eqnarray}
ds^2 = a^2(\tau)\{-(1+2\phi)d\tau^2 + 2B_i^{(s)} d\tau dx^i   \nonumber  \\
  \hspace{20mm} +[(1-2\psi)\delta_{ij} +2E_{ij}^{(s)}]dx^i dx^j\} \,,	
		\label{genmetric}
\end{eqnarray}
where $B_i^{(s)}$ and $E_{ij}^{(s)}$ can be written as
\begin{eqnarray}
B_i^{(s)} &=&-\frac{ik_i}{k}B \,, \\ 
E_{ij}^{(s)} &=& \left(-\frac{k_ik_j}{k^2}+\frac{1}{3}\delta_{ij}\right)E \,,
\end{eqnarray}
where $\phi,B,\psi,E$ are arbitrary scalar functions which describe
the perturbations on a Friedmann-Robertson-Walker (FRW) background.
Performing a first order coordinate change, $\tilde{\tau}=\tau+\xi^0$,
$\tilde{x^i}=x^i+\xi^i$, the scalar metric variables in the new gauge
(denoted by a tilde) are then given by
\begin{eqnarray}
\tilde{\phi} &= &\phi-\xi^{0'}-h\xi^0 \,, \label{metrictransform1} \\
\tilde{B} &=&B-\xi'+k\xi^0 \,, \\
\tilde{\psi} &=&\psi+h\xi^0 \,,	\label{psitrans}  \\
\tilde{E} &=&E-k\xi \,,  \label{metrictransform4}
\end{eqnarray}
where the scalar part of $\xi^i(\tau,x^i)$ is defined as
$\xi^{i(s)}=-\frac{ik^i}{k}\xi$, primes denote derivatives with
respect to conformal time $\tau$, and $h=a'/a=aH$.  The density contrast and the
velocity perturbation transform as
\begin{eqnarray}
\label{fluidtransform}
\tilde{\delta} &=&\delta+3h(1+w)\xi^0 \,, \\
\label{vtransform}
\tilde{V} &=&V+\xi' \, ,			
\end{eqnarray}
where $V$ is related to the scalar part of the 3-velocity vector,
$v^{i(s)}=-\frac{ik^i}{k}V$.

Ref.~\cite{Green:2005fa} calculated the evolution of the density and
velocity perturbations in the conformal Newtonian gauge (which has
$B_N=E_N=0$). 

They found that during radiation domination ($w=1/3$) for a fluid with
vanishing anisotropic stress (and hence $\phi_N=\psi_N$), the
remaining perturbations evolve according to
\begin{eqnarray}
\phi_N &=&\frac{j_1(\kappa)}{\sqrt{3}\kappa}C  \,, \label{cnphi}  \\
\delta_N &= &\frac{2}{\sqrt{3}}\left(2\frac{j_1(\kappa)}{\kappa}-j_0(\kappa)-\kappa j_1(\kappa)\right)C \,, \label{cndelta}  \\
V_N &= &\left(j_1(\kappa)-\frac{\kappa}{2}j_0(\kappa)\right)C \,,  \label{cnv}
\end{eqnarray}
where $\kappa=k/\sqrt{3} aH$ and $C$ is a
normalisation constant.

For the PBH abundance calculation we need the density perturbation in
the total matter gauge (T), which is defined by $B_T+ V_T =0$, and $E_T=0$.
%
Since the comoving curvature perturbation, $\mathcal{R}$, is identical
to the curvature perturbation in the total matter gauge, $\psi_T$ (see
e.g.~\cite{Malik:2008im}), we get using
eqs.~(\ref{metrictransform1}-\ref{metrictransform4}),
%
%
\begin{equation}
\mathcal{R}=\phi_N-\frac{h}{k}V_N\,.
\end{equation}

Similarly, the density contrast in the total matter gauge (during
radiation domination) is
\begin{equation}
\label{dV1}
\delta_T=\delta_N-4\frac{h}{k}V_N\,.
\end{equation}

Using the solutions eqs.~(\ref{cnphi}) and (\ref{cnv}) above we get
\begin{equation}
\mathcal{R}=\frac{C}{2\sqrt{3}}j_0(\kappa)\,,
\end{equation}
which reduces in the small $\kappa$ (large scale) limit to
$C=2\sqrt{3}\mathcal{R}$, allowing us to replace the normalisation
constant $C$ with the large scale limit of $\mathcal{R}$.
Equation (\ref{dV1}) then becomes, using eqs.~(\ref{cndelta}) and (\ref{cnv}),
\begin{equation}
\delta_T= -\frac{4}{\sqrt{3}} \left(\frac{k}{aH}\right) 
j_{1}(k/\sqrt{3}aH) \mathcal{R} \,.
\end{equation}




\providecommand{\href}[2]{#2}\begingroup\raggedright\endgroup

\end{document}